\documentclass[amsmath,amssymb,twocolumn,nofootinbib]{revtex4-1}
\usepackage{graphicx}
\usepackage{epsfig}
\usepackage{xcolor}
\usepackage[normalem]{ulem}

\begin{document}
\title{Regular black holes, universes without singularities, and phantom-scalar field transitions}
\author{Leonardo Chataignier}
\email{leonardo.chataignier@unibo.it}
\affiliation{Dipartimento di Fisica e Astronomia, Universit\`{a} di Bologna,
via Irnerio 46, 40126 Bologna, Italy\\I.N.F.N., Sezione di Bologna, I.S. FLAG, viale B. Pichat 6/2, 40127 Bologna, Italy}
\author{Alexander~Yu.~Kamenshchik}
\email{kamenshchik@bo.infn.it}
\affiliation{Dipartimento di Fisica e Astronomia, Universit\`{a} di Bologna,
via Irnerio 46, 40126 Bologna, Italy\\I.N.F.N., Sezione di Bologna, I.S. FLAG, viale B. Pichat 6/2, 40127 Bologna, Italy}
\author{Alessandro Tronconi}
\email{tronconi@bo.infn.it}
\affiliation{Dipartimento di Fisica e Astronomia, Universit\`{a} di Bologna,
via Irnerio 46, 40126 Bologna, Italy\\I.N.F.N., Sezione di Bologna, I.S. FLAG, viale B. Pichat 6/2, 40127 Bologna, Italy}
\author{Giovanni Venturi}
\email{giovanni.venturi@bo.infn.it}
\affiliation{Dipartimento di Fisica e Astronomia, Universit\`{a} di Bologna,
via Irnerio 46, 40126 Bologna, Italy\\I.N.F.N., Sezione di Bologna, I.S. FLAG, viale B. Pichat 6/2, 40127 Bologna, Italy}
\begin{abstract}
We consider a procedure of elimination of cosmological singularities similar to that suggested in the recent paper by Simpson and Visser for the construction of regular black holes. It is shown that by imposing a non-singular cosmological evolution with a bounce in a flat Friedmann universe filled with a minimally coupled scalar field, we obtain a transition between the standard scalar field and its phantom counterpart. In this case, the potential of the scalar field has a non-analyticity of the cusp type. This result is also readily reproduced in the case of an anisotropic Bianchi I universe.  We have also found a spherically symmetric static solution of the Einstein equations, free of singularities and sustained by a scalar field. 
\end{abstract}
\maketitle
\section{Introduction}
The study of non-singular black holes has a rather long history \cite{Bardeen} (for recent reviews, see \cite{Spallucci,Sebastiani}). In the recent paper by Simpson and Visser \cite{Simpson-Visser}, it was shown that one may obtain a singularity-free spacetime from the Schwarzschild black hole by a  simple substitution of the radial coordinate $r$ by the function $r(u) = \sqrt{u^2+b^2}$, where $u$ is a new coordinate and $b$ is the regularization parameter. Thus, the new metric obtained from the Schwarzschild metric has the following form:
\begin{widetext}
\begin{equation}
ds^2 = \left(1-\frac{2m}{\sqrt{u^2+b^2}}\right)dt^2-\left(1-\frac{2m}{\sqrt{u^2+b^2}}\right)^{-1}du^2-(u^2+b^2)(d\theta^2+\sin^2\theta d\varphi^2),
\label{S-V}
\end{equation} 
\end{widetext}
where the singularity at $r=0$ is replaced by a regular minimum of $r(u)$ at $u=0$, a sphere of radius $b$. If $b > 2m$, the formula (1) represents a wormhole with a throat 
at $u=0$; if $b <2m$, one has a black hole with two horizons at $u=\pm\sqrt{4m^2-b^2}$; and if $b=2m$, we see an extremal black hole with a single horizon at $u=0$.

In the black hole case, the hypersurface $u=0$ is not a throat since $u$ is a temporal coordinate there, and $u=0$ corresponds to a bounce in one of the two scale factors of the Kantowski-Sachs universe in the inner region of the black hole. This phenomenon was called the black bounce in \cite{Simpson-Visser}. Later, a similar mechanism was used to regularize a Vaidya spacetime \cite{Vaidya}, charged black-bounce spacetimes \cite{charged}, and Kerr black holes \cite{Kerr}. The field sources for the Simpson-Visser spacetimes were considered in the article \cite{Bron}. Various black-bounce solutions of general relativity were analyzed in \cite{Canate:2022}.

Moreover, in the paper by Bronnikov \cite{Bronnikov}, a regularization of the static spherically symmetric configuration in the presence of a scalar field was considered. The solution was the regularized version of the Fisher solution \cite{Fisher}, and it could be seen as an exact solution of general relativity in the presence of a self-interacting scalar field and of a magnetic field (it is important to note that the Fisher solution was rediscovered many times and studied in different contexts \cite{Fisher1,Fisher2,Fisher3,Fisher4,Fisher5,Fisher6,Fisher7,we}).  
An interesting transformation between the phantom scalar field and the standard scalar field was obtained. 

In this paper, we investigate some effects that result from the application of the Simpson-Visser-like prescription directly to cosmology. It is well known that there is a deep connection between the static solutions of the Einstein equations and their cosmological solutions. We have already mentioned that the interior part of the Schwarzschild black holes is nothing but the Kantowski-Sachs universe. It is less known that there is a duality between spherically symmetric static solutions of the Einstein equations in the presence of a scalar field and the Kantowski-Sachs universe with hyperbolic spatial sections. Inversely, the Kantowski-Sachs solutions with spherical spatial sections correspond to static solutions with hyperbolic symmetry. This phenomenon was noticed in the article \cite{we} and further investigated in articles \cite{we1,we2,we3}. This duality is useful because, for some problems, working with time-dependent metrics is more convenient, while static metrics are preferred for other applications.  

Here, it is necessary to mention that the attempts to construct the non-singular cosmological models, i.e., models describing a universe without an initial singularity, have a very long history and are still of considerable interest. The Einstein static universe \cite{Einstein} was the first cosmological model based on general relativity. Another historically important cosmological model without singularities is the steady-state universe \cite{steady,steady1}, which for a certain period of time was a strong competitor of the Big Bang theory. More modern models where an initial singularity was absent were developed in the framework of string cosmology \cite{string,string1,string2} and loop quantum cosmology \cite{lqc}. Quite a few recent works devoted to non-singular cosmology \cite{modern,modern1,modern2,modern3} used modified gravity theories such as Galileon gravity \cite{Galileon} and Horndenski gravity \cite{Horn}.  A distinguishing feature of these models is a very special role played by a scalar field: arbitrary functions of the scalar field appear in the Lagrangian, but the equations of motion do not contain terms with derivatives higher than that of second order. It is also worth mentioning that the bounces, and hence the absence of initial or final singularities, can be achieved on a simple closed Friedmann model filled with a minimally coupled scalar field \cite{bounce,bounce1,bounce2,bounce3,bounce4,bounce5,bounce6}. In this case, there exist both singular and non-singular solutions of the equations of motion, and the study of the structure of the set of non-singular trajectories reveals some interesting features.

The construction of exact solutions to the Einstein equations is an important task because it can help in the analysis of various aspects of singularities (including their possible avoidance), and in the discovery of other interesting features. One can search for exact solutions, e.g., by fixing symmetries of the spacetime and properties of the matter fields, such as their equations of state or Lagrangian density. However, one can also work in another way: one can write down a metric for spacetime, substitute it into the Einstein equations and observe which kind of matter emerges on the right-hand side of these equations. This approach can be justified when it leads to an interesting phenomenology of spacetime and matter. Precisely this method was applied in the paper by Simpson and Visser \cite{Simpson-Visser} and in subsequent articles. Namely, in the well-known solutions of the Einstein equations, a simple substitution $r \rightarrow \sqrt{r^2+b^2}$ is considered, and it transforms these singular solutions into regular ones. Evidently, a similar method can be applied to the singular cosmological solutions. The analogy is direct: one can make the substitution $t \rightarrow \sqrt{t^2+b^2}$ and obtain the non-singular cosmological solutions. 

Here, we show that already in a flat Friedmann model, using the Simpson-Visser-like regularization, one can observe both a bounce (which replaces the cosmological singularity) and the phantom-scalar transition. We shall also briefly discuss the connection of these effects with similar results discovered in a slightly different cosmological context \cite{we4,we5}. 

Then, taking into account the important role played by scalar fields in modern gravity and cosmology, and inspired by the article \cite{Bronnikov}, we set out to find a regular Fisher-like static spherically symmetric solution obtained solely from a scalar field
(and not from the combination of a scalar field and an electromagnetic field as in \cite{Bronnikov}). To find this solution, we make the substitution of the type $r \rightarrow \sqrt{r^2+b^2}$ only in the scale factor that multiplies the metric of the two-dimensional sphere, while the other two metric coefficients maintain an arbitrary form. The solution that is obtained is simple and possesses interesting features, which we consider in some detail.

The structure of the paper is the following: in the next section, we consider a flat Friedmann universe filled with a scalar field, and we show that the simple regularization of the metric implies the phantom-scalar transition with a cusp type of non-analyticity in the scalar field potential; in the third section we show that these results can also be reproduced in an anisotropic Bianchi I universe; the fourth section is devoted to spherically symmetric static regular geometries sustained by a scalar field. The last section contains our concluding remarks.

\section{Flat Friedmann model with a scalar field}
Let us consider a flat Friedmann universe filled with a massless scalar field. The exact solution for the metric and for the scalar field in this model is well known and has the following form:
\begin{equation}
ds^2=dt^2-t^{\frac23}(dx_1^2+dx_2^2+dx_3^2),
\label{Fried}
\end{equation}
\begin{equation}
\dot{\phi} = \sqrt{\frac23}\frac{1}{t},
\label{KG}
\end{equation}
where a dot denotes the differentiation with respect to the cosmic time $t$ and the normalization is chosen so as to simplify the form of the equations.

Let us now construct the regularized metric following the recipe of article \cite{Simpson-Visser}:
\begin{equation}
ds^2 = dt^2 -(t^2+b^2)^{\frac13}(dx_1^2+dx_2^2+dx_3^2).
\label{Fried1}
\end{equation}
A straightforward calculation yields the expressions for the Ricci tensor components:
\begin{equation}
R^0_0 = \frac{2t^2-3b^2}{3(t^2+b^2)^4},
\label{Ricci}
\end{equation}
\begin{equation}
R_1^1=R_2^2=R_3^3=-\frac{b^2}{3(t^2+b^2)^2}.
\label{Ricci1}
\end{equation}
The Ricci scalar is 
\begin{equation}
R = \frac{2t^2-6b^2}{3(t^2+b^2)^2}.
\label{Ricci2}
\end{equation}
The Friedmann equations immediately give us the expressions for the energy density and for the isotropic pressure of matter
\begin{equation}
\rho=\frac{t^2}{3(t^2+b^2)^2},
\label{energy}
\end{equation}
\begin{equation}
p = \frac{t^2-2b^2}{3(t^2+b^2)^2}.
\label{pressure}
\end{equation}
As for the unregularized case, let us suppose that the universe is filled with a spatially homogeneous scalar field with a potential $V(\phi)$. Then
\begin{equation}
\rho = \frac12\dot{\phi}^2+V(\phi)
\label{energy1}
\end{equation}
and 
\begin{equation}
p = \frac12\dot{\phi}^2-V(\phi).
\label{pressure1}
\end{equation}
Comparing Eqs. (\ref{energy}) and (\ref{pressure}) with Eqs. (\ref{energy1}) and (\ref{pressure1}) we obtain
\begin{equation}
\dot{\phi} = \pm\sqrt{\frac23}\frac{\sqrt{t^2-b^2}}{t^2+b^2},
\label{scalar}
\end{equation}
and 
\begin{equation}
V= \frac{b^2}{3(t^2+b^2)^2}.
\label{poten}
\end{equation}
Equation \eqref{scalar} can be integrated, and we can find the field $\phi$ as a function of time $t$. However, we are not able to invert the result or find $t$ as an explicit function of $\phi$, and thus we cannot use Eq. \eqref{poten} to find the explicit form of the potential in terms of the scalar field. Nevertheless, Eqs. \eqref{scalar} and \eqref{poten} contain rather interesting information. We see that the expression \eqref{scalar} is well defined if $|t| \geq |b|$. What happens if $|t| < |b|$? The kinetic energy of $\phi$ changes sign and the standard scalar field transitions to a phantom scalar field. Thus, we observe an effect analogous to that described in \cite{Bronnikov}. 

As far as the form of the potential is concerned, we can study its behavior in the vicinity of $t = b$. Let us define
\begin{equation}
t = b+\tau,
\label{time}
\end{equation}
where $\tau$ is small. Then, from Eq. \eqref{scalar} with the choice of an overall plus sign (which is not essential), we obtain
\begin{equation}
\frac{d\phi}{d\tau} = \frac{\sqrt{\tau}}{\sqrt{3b^3}}
\label{scalar1}
\end{equation}
and 
\begin{equation}
\phi(\tau) = \phi_0+\frac{2\tau^{3/2}}{3\sqrt{3b^3}},
\label{scalar2}
\end{equation}
where $\phi_0$ is an integration constant. 
Hence,
\begin{equation}
\tau = 3b\left(\frac{\phi-\phi_0}{2}\right)^{\frac23}.
\label{time1}
\end{equation}
Substituting the expressions \eqref{time} and \eqref{time1} into Eq. \eqref{poten}, we obtain the formula describing the behavior of the potential of the scalar field in the vicinity of the critical point:
 \begin{equation}
 V(\phi) = \frac{1}{3b^2\left[\left(1+3\left(\frac{\phi-\phi_0}{2}\right)^{\frac23}\right)^2+1\right]^2}.
 \label{poten1}
 \end{equation}
By keeping only the leading terms, we can rewrite the above expression as follows:
\begin{equation}
V(\phi) = \frac{1}{12b^2}\left[1-6\left(\frac{\phi-\phi_0}{2}\right)^{\frac23}\right].
\label{poten2}
\end{equation}
The distinguishing feature of Eq. \eqref{poten2} is the presence of a non-analyticity of the cusp type, which is responsible for the transition from the standard scalar field to its phantom counterpart and vice versa.   

We can also consider a slightly more general model: a flat Friedmann model with the metric 
\begin{equation}
ds^2=dt^2-t^{2\alpha}(dx_1^2+dx_2^2+dx_3^2) \,,
\label{Fried2}
\end{equation}
which arises in a universe filled with a perfect fluid with the equation-of-state parameter
\begin{equation}
w = \frac{2-3\alpha}{3\alpha}.
\label{eq-state}
\end{equation}
It is well known that this is a particular solution to the equations of motion for the flat Friedmann model with a minimally coupled scalar field with an exponential potential (see, e.g., \cite{we6} and references therein). In order to avoid the cosmological singularity, we modify the metric \eqref{Fried2} as follows
\begin{equation} 
ds^2=dt^2-(t^2+b^2)^{\alpha}dl^2.
\label{Fried3}
\end{equation}
The components of the Ricci tensor now read
\begin{equation}
R_0^0 = -\frac{3\alpha(b^2+(\alpha-1)t^2)}{(t^2+b^2)^2},
\label{Ricci3}
\end{equation}
\begin{equation}
R_1^1=R_2^2=R_3^3=-\frac{\alpha(b^2+(3\alpha-1)t^2)}{(t^2+b^2)^2}.
\label{Ricci4}
\end{equation}
The Ricci scalar is 
\begin{equation}
R=-\frac{6\alpha(b^2+(2\alpha-1)t^2)}{(t^2+b^2)^2}.
\label{Ricci5}
\end{equation}
Now we can find the expressions for the energy density 
\begin{equation}
\rho=\frac{3\alpha^2t^2}{(t^2+b^2)^2} ,
\label{energy2}
\end{equation}
and the pressure
\begin{equation}
p=-\frac{\alpha(2b^2+(3\alpha-2)t^2)}{(t^2+b^2)^2}.
\label{pressure2}
\end{equation}
The expressions for the potential and the time derivative of the scalar field realizing the evolution \eqref{Fried3} are
\begin{equation}
V(\phi) = \frac{\alpha(b^2+(3\alpha-1)t^2)}{(t^2+b^2)^2},
\label{poten3}
\end{equation}
\begin{equation}
\dot{\phi}^2=\frac{2\alpha(t^2-b^2)}{(t^2+b^2)^2}.
\label{scalar2}
\end{equation}
We note that, when the regularizing parameter $b=0$, one can easily obtain from Eqs. \eqref{poten3} and \eqref{scalar2} the known expression for the exponential potential:
\begin{equation}
V(\phi) = \alpha(3\alpha-1)\exp\left(-\sqrt{\frac{2}{\alpha}}(\phi-\phi_0)\right).
\label{poten4}
\end{equation}
Alternatively, if $b > 0$, we can see that, just as in the case considered above (where a particular value $\alpha = \frac13$ was chosen), the transition from the standard scalar field to the phantom (or vice versa) takes place. Now, we can again consider the vicinity of the instant $t=b$ [see Eq. \eqref{time}]. Proceeding in a similar way to what was shown above, we obtain the following expression for the behavior of the potential in the vicinity of the cusp:
\begin{equation}
V(\phi) = \frac{\alpha}{4b^2}\left[3\alpha-\frac{2\cdot3^{2/3}}{\alpha^{1/3}}\left(\frac{\phi-\phi_0}{2}\right)^{2/3}\right].
\label{poten5}
\end{equation}
This expression has the same non-analyticity [$\sim (\phi-\phi_0)^{2/3}$] as that seen in the expression \eqref{poten2}, and, when $\alpha = \frac13$, these expressions coincide.

\section{Scalar field - phantom transitions in a Bianchi I universe}
Let us consider a Bianchi I universe with the metric 
\begin{equation}
ds^2 = dt^2-(a_1^2(t)dx_1^2+a_2^2(t)dx_2^2+a_3^2(t)dx_3^2).
\label{Bianchi}
\end{equation}
It is convenient to introduce the following variables:
\begin{eqnarray}
&&a_1(t) = A(t)e^{\beta_1(t)},\nonumber \\
&&a_2(t) = A(t)e^{\beta_2(t)},\nonumber \\
&&a_3(t) = A(t)e^{\beta_3(t)},
\label{Bianchi1}
\end{eqnarray}
where the anisotropic factors $\beta_i$ satisfy the identity
\begin{equation}
\beta_1+\beta_2+\beta_3 = 0.
\label{Bianchi2}
\end{equation}
The Ricci tensor components and the Ricci scalar have the following form:
\begin{eqnarray}
&&R_0^0 = -3\frac{\ddot{A}}{A}-\sum_{i=1}^3\dot{\beta}_i^2,\nonumber \\
&&R_1^1 = -\left(\frac{\ddot{A}}{A}+2\frac{\dot{A}^2}{A^2}+3\frac{\dot{A}}{A}\dot{\beta}_1-\ddot{\beta}_1\right), \nonumber \\
&&R_2^2 = -\left(\frac{\ddot{A}}{A}+2\frac{\dot{A}^2}{A^2}+3\frac{\dot{A}}{A}\dot{\beta}_2-\ddot{\beta}_2\right), \nonumber \\ 
&&R_3^3 = -\left(\frac{\ddot{A}}{A}+2\frac{\dot{A}^2}{A^2}+3\frac{\dot{A}}{A}\dot{\beta}_3-\ddot{\beta}_3\right), \nonumber \\
&&R = -\left(6\frac{\ddot{A}}{A}+6\frac{\dot{A}^2}{A^2}+\sum_{i=1}^3\dot{\beta}_i^2\right).
\label{new-param2} 
\end{eqnarray}
Now, if we have an empty space or a space filled with matter with an isotropic pressure, we have 
\begin{equation}
R_1^1=R_2^2=R_3^3,
\label{Bianchi3}  
\end{equation}
and hence
\begin{equation}
 R_2^2+R_3^3-2R_1^1 = 0.
 \label{Bianchi4}
 \end{equation}
Substituting the expressions \eqref{new-param2} into Eq. \eqref{Bianchi4} and using the relation \eqref{Bianchi2}, we obtain 
 \begin{equation}
 \ddot{\beta}_1+3\frac{\dot{A}}{A}\dot{\beta}_1 = 0. 
 \label{new-param3}
 \end{equation} 
 Integrating this equation, we obtain 
 \begin{equation}
 \dot{\beta}_1 = \frac{\beta_{10}}{A^3},
 \label{new-param4}
 \end{equation}
 where $\beta_{10}$ is an integration constant. Similarly, we obtain 
 \begin{equation}
 \dot{\beta}_2 = \frac{\beta_{20}}{A^3},
 \label{new-param5}
 \end{equation}
 \begin{equation}
 \dot{\beta}_3 = \frac{\beta_{30}}{A^3}.
 \label{new-param6}
 \end{equation}
If we now substitute Eqs. (\ref{new-param4})--(\ref{new-param6}) into the expressions for the components of the Ricci curvature and of the Ricci scalar, we find
 \begin{equation}
 R_0^0-\frac12R = 3\frac{\dot{A}^2}{A^2}-\frac12\frac{\bar{\beta}^2}{A^6},
\label{Fried-gen}
\end{equation}
where 
\begin{equation}
\bar{\beta}^2 \equiv \sum_{i=1}^3\beta_{i0}^2.
 \label{gen-Fried1}
 \end{equation}
If the universe is empty, we can equate the right-hand side of Eq. \eqref{Fried-gen} to zero, from which we can find the scale factor $A$. It is also possible to compute the explicit expressions for the anisotropy factors by integrating Eqs. \eqref{new-param4}--\eqref{new-param6}, and the result is the famous Kasner solution \cite{Kasner}. By considering the universe filled with dust or other perfect fluids, we obtain the Heckmann-Schucking solution \cite{Heck-Schuck} or its generalizations \cite{Jacobs,Khalat}. 

Here, we use instead the same prescription, which was implemented in the preceding section, and suppose that  
\begin{equation}
A(t) = (t^2+b^2)^{\frac16}.
\label{Bianchi5}
\end{equation}
When the regularizing parameter $b=0$, we obtain the Kasner solution. 
Now, if $b > 0$, the energy density of the matter filling the universe is 
\begin{equation}
\rho = \frac{t^2}{3(t^2+b^2)^2}-\frac{\bar{\beta}^2}{2(t^2+b^2)},
\label{en-Bian}
\end{equation}
while the pressure is
\begin{equation}
p = \frac{t^2-2b^2}{3(t^2+b^2)^2}-\frac{\bar{\beta}^2}{2(t^2+b^2)}.
\label{pres-Bian}
\end{equation}
If the universe is filled with a minimally coupled scalar field, we can find the expressions for its potential and the kinetic term using Eqs.  \eqref{energy1} and \eqref{pressure1} 
from the preceding section. We obtain 
\begin{equation}
\dot{\phi}^2 = \frac{t^2(2-3\bar{\beta}^2)-b^2(2+3\bar{\beta}^2)}{3(t^2+b^2)^2}. 
\label{Bian-scal}
\end{equation}
\begin{equation}
V = \frac{b^2}{3(t^2+b^2)^2}.
\label{Bian-poten}
\end{equation}
Note that the expression for the potential as a function of time coincides with that obtained in the preceding section for the case of the flat Friedmann universe \eqref{poten}. The expression for the time derivative of the scalar field \eqref{Bian-scal} is modified by the presence of the anisotropy. One can see that the phantom-scalar transition occurs at 
\begin{equation}
|t| = b\sqrt{\frac{2+3\bar{\beta}^2}{2-3\bar{\beta}^2}}.
\label{time-Bian}
\end{equation}
We also note that the presence of the scalar field imposes the restriction on the value of anisotropy:
\begin{equation}
\bar{\beta}^2 \leq \frac23.
\label{time-Bian1}
\end{equation}
This effect was observed for the Bianchi I universe filled with a massless scalar field years ago \cite{Bel-Khal}. Just as in the case of the flat Friedmann model, the potential providing the ``(de)-phantomization'' of the scalar field has a cusp.

\section{Spherically symmetric static regular geometries sustained by a scalar field}

As we have already mentioned in the Introduction, the article \cite{Bronnikov}  considered regularized Fisher-type solutions, where the role of matter was played by the scalar field and by the electromagnetic field. Here, we wish to construct a spherically symmetric static spacetime filled exclusively with the scalar field. As in the preceding sections, we consider a scalar field minimally coupled with gravity that can undergo a phantom-scalar transition.

We look for the solution in the following form:
\begin{equation}
ds^2=A(r)dt^2-B(r)dr^2 - (r^2+b^2)(d\theta^2+\sin^2\theta d\varphi^2).
\label{stat}
\end{equation} 
The radial coordinate $r$, the redshift function $A(r)$ and the profile function $B(r)$ are free, and the function that multiplies the two-dimensional sphere metric is chosen as $r^2+b^2$. (One can note that this choice is different from the one usually used for the construction of the Fisher-type solutions.) 

We can calculate the components of the Ricci tensor for this metric:
\begin{equation}
R_0^0 = \frac{A''}{2AB}-\frac{A'^2}{4A^2B}-\frac{A'B'}{4AB^2}+\frac{A'r}{AB(r^2+b^2)},
\label{stat-Ricci}
\end{equation} 
\begin{equation}
R_r^r = \frac{A''}{2AB}-\frac{A'^2}{4A^2B}-\frac{A'B'}{4AB^2}-\frac{B'r}{B^2(r^2+b^2)}+\frac{2b^2}{B(r^2+b^2)^2},
\label{stat-Ricci1}
\end{equation}
\begin{eqnarray}
&&R_{\theta}^{\theta}=R_{\varphi}^{\varphi}=\frac{1}{B(r^2+b^2)}-\frac{B'r}{2B^2(r^2+b^2)}\nonumber \\
&&+\frac{A'r}{2AB(r^2+b^2)}-\frac{1}{r^2+b^2},
\label{stat-Ricci2}
\end{eqnarray}
where the prime denotes the derivative with respect to $r$.

Let us now suppose that our spacetime is filled with a scalar field that depends solely on the radial coordinate $r$. Its Lagrangian is then
\begin{equation}
L = \frac{1}{2}g^{\mu\nu}\phi_{,\mu}\phi_{,\nu}-V(\phi) = -\frac{1}{2B}\phi'^2-V(\phi). 
\label{stat-Lagrange}
\end{equation}
The components of the energy-momentum tensor are
\begin{equation}
T_0^0 = \frac{1}{2B}\phi'^2+V(\phi),
\label{stat-en-mom}
\end{equation}
\begin{equation}
T_r^r = -\frac{1}{2B}\phi'^2+V(\phi),
\label{stat-en-mom1}
\end{equation}
\begin{equation}
T_{\theta}^{\theta} = T_{\varphi}^{\varphi}=\frac{1}{2B}\phi'^2+V(\phi).
\label{stat-en-mom2}
\end{equation}
We see that 
\begin{equation}
T_0^0=T_{\theta}^{\theta}=T_{\varphi}^{\varphi},
\label{stat-equality}
\end{equation}
and hence the corresponding components of the Einstein tensor and of the Ricci tensor should also coincide:
\begin{equation}
R_0^0=R_{\theta}^{\theta}=R_{\varphi}^{\varphi}.
\label{stat-equality1}
\end{equation}
Substituting the expressions \eqref{stat-Ricci} and \eqref{stat-Ricci2} into Eq. \eqref{stat-equality1}, we obtain the following condition:
\begin{eqnarray}
&&\frac{A''}{2AB}-\frac{A'^2}{4A^2B}-\frac{A'B'}{4AB^2}+\frac{A'r}{2AB(r^2+b^2)}\nonumber \\
&&-\frac{1}{B(r^2+b^2)}+\frac{B'r}{2B^2(r^2+b^2)}+\frac{1}{r^2+b^2} = 0.
\label{stat-condition}
\end{eqnarray}
The above equation imposes a restriction on the choice of the functions $A$ and $B$ but does not determine them completely. A reasonable solution can be obtained with an additional hypothesis concerning the form of these functions. One of the simplest possible choices is the ``Schwarzschild-like'' condition 
\begin{equation}
AB = 1.
\label{stat-condition1}
\end{equation}
Using this condition, we can rewrite Eq. \eqref{stat-condition} as an ordinary linear second-order differential equation for the function $A$:
\begin{equation}
A'' - \frac{2A}{r^2+b^2}+\frac{2}{r^2+b^2} = 0.
\label{stat-condition2}
\end{equation}
It is straightforward to find the general solution of Eq. \eqref{stat-condition2}. A particular solution of the inhomogeneous equation is
\begin{equation}
A_0 = 1.
\label{stat-part}
\end{equation}
An obvious solution of the homogenous equation is 
\begin{equation}
A_1 = r^2+b^2.
\label{stat-gen}
\end{equation} 
To find the second independent solution of the homogeneous equation, we can use the Wronskian relation:
\begin{equation}
A_2'A_1-A_1'A_2 = 1,
\label{Wron}
\end{equation}
which immediately gives
\begin{equation}
A_2'-\frac{2r}{r^2+b^2}A_2-\frac{1}{r^2+b^2} = 0.
\label{Wron1}
\end{equation} 
Looking for the solution of this equation in the form 
\begin{equation}
A_2 = (r^2+b^2)f,
\label{Wron2}
\end{equation}
we obtain
\begin{equation}
f' = \frac{1}{(r^2+b^2)^2}.
\label{Wron3}
\end{equation}
Integrating Eq. \eqref{Wron3}, we obtain 
\begin{equation}
f = \frac{1}{2b^3}{\rm arctan}\frac{r}{b}+\frac{r}{2b^2(r^2+b^2)} + const.
\label{Wron4}
\end{equation}
 Finally, the general solution of Eq. \eqref{stat-condition2} is 
 \begin{equation}
 A = 1+c_1(r^2+b^2) +\frac{c_2}{2b^3}\left[(r^2+b^2){\rm arctan}\frac{r}{b}+br\right],
 \label{stat-gen1}
 \end{equation} 
where $c_1$ and $c_2$ are arbitrary constants. 

It is reasonable to require that the general solution \eqref{stat-gen1} is well defined at the limit $b \rightarrow 0$. We use the expansion of the arctan function when its argument tends to infinity,
\begin{equation}
{\rm arctan}\frac{r}{b} = \frac{\pi}{2}-\frac{b}{r}+\frac{b^3}{3r^3}-\frac{b^5}{5r^5}+\cdots.
\label{arctan}
\end{equation}
Substituting the expansion \eqref{arctan} into the expression \eqref{stat-gen1}, we find that the condition for the regularity of this expression at $b \rightarrow 0$ is 
\begin{equation}
c_1 = -\frac{c_2\pi}{4b^3}.
\label{stat-condition3}
\end{equation}
We note that the expansion \eqref{arctan} is also valid when the parameter $b$ is fixed while the radius $r$ tends to infinity. In this limit, and substituting \eqref{stat-condition3}, we have  
\begin{equation}
A = 1 - \frac{c_2}{3r}+\frac{c_2b^2}{15r^3}+ \cdots.
\label{stat-condition4}
\end{equation}
It is convenient to introduce the notation $r_0 = \frac{c_2}{3}$ when at $r \rightarrow \infty$ the expression \eqref{stat-condition4} has a ``Schwarzschild-like'' form
\begin{equation}
A = 1- \frac{r_0}{r} + \frac{b^2r_0}{5r^3}+\ldots \ .
\label{stat-condition5}
\end{equation}
Finally, the expression for $A(r)$ that we have obtained is 
\begin{equation}
A = 1- \frac{3\pi r_0}{4b^3}(r^2+b^2)+\frac{3r_0}{2b^3}\left[(r^2+b^2){\rm arctan}\frac{r}{b}+br\right].
\label{stat-exact}
\end{equation}
This geometry does not have any singularity at $r=0$. 

Now, we would like to understand in which cases the metric \eqref{stat-exact} describes a regular black hole or a wormhole. We write down the derivatives of the function $A$:
\begin{equation}
A' = -\frac{3\pi r_0r}{2b^3}+\frac{3r_0r}{b^3}{\rm arctan}\frac{r}{b}+\frac{3r_0}{b^2},
\label{deriv}
\end{equation}
\begin{equation}
A'' = -\frac{3\pi r_0}{2b^3}+\frac{3r_0}{b^3}{\rm arctan}\frac{r}{b}+\frac{3r_0r}{b^2(r^2+b^2)},
\label{deriv1}
\end{equation}
\begin{equation}
A'''= \frac{6r_0}{(r^2+b^2)^2}.
\label{deriv2}
\end{equation}
We see that the third derivative $A'''$ is always positive. The second derivative at $r=0$ is 
\begin{equation}
A''(0) = -\frac{3\pi r_0}{2b^3} < 0.
\label{deriv3}
\end{equation}
Using the expansion \eqref{arctan}, we find that at $r \rightarrow \infty$ 
\begin{equation}
A''(r) = -\frac{2r_0}{r^3} + \ldots \ ;
\label{deriv4}
\end{equation}
i.e., it tends to zero remaining negative. This means that the second derivative $A''$ is always negative. The first derivative $A'$ at $r=0$ is equal to 
\begin{equation}
A'(0) = \frac{3r_0}{b^2} \ , 
\label{deriv5}
\end{equation}
and it is positive. At $t \rightarrow \infty$, it behaves like 
\begin{equation}
A' = \frac{r_0}{r^2}+\ldots \ ;
\label{deriv6}
\end{equation} 
i.e., it tends to zero, remaining positive, and hence it is always positive. The value of the function $A$ at $r=0$ can be either negative or positive depending on the relation between the parameters $b$ and $r_0$. If $r \rightarrow \infty$, the function $A$ tends to $1$ as follows from Eq. \eqref{stat-condition5}. Thus, if it is negative at $r=0$, it changes sign and this situation corresponds to a regular black hole, while in the opposite case it describes a wormhole with a throat at $r=0$. Finally,  if 
\begin{equation}
b \geq \frac{3\pi r_0}{4} \ ,
\label{worm}
\end{equation}
one has a wormhole. In the opposite case, we have a regular black hole with a black bounce. We also note that the geometry presented above in the case of the regular black hole has only one horizon. Indeed, as it is clear from the analysis of the function $A(r)$ and its derivatives, it crosses the value $A=0$ only once. 

We can now connect this spacetime geometry with a dynamical scalar field using the Einstein equations
\begin{equation}
G_{\mu}^{\nu} = T_{\mu}^{\nu},
\label{Ein}
\end{equation}
where we have chosen convenient units. Combining these equations with Eqs. \eqref{stat-en-mom} and \eqref{stat-en-mom1}, we see that 
\begin{equation}
\phi'^2 = B(T_0^0-T_r^r)=B(G_0^0-G_r^r)=B(R_0^0-R_r^r).
\label{stat-scal}
\end{equation}
Using the explicit expressions for the components of the Ricci tensor \eqref{stat-Ricci} and \eqref{stat-Ricci1} together with the condition \eqref{stat-condition1}, we obtain
\begin{equation}
\phi'^2 = -\frac{b^2}{(r^2+b^2)^2}.
\label{stat-scal1}
\end{equation} 
The negative definiteness of the right-hand side of Eq. \eqref{stat-scal1} indicates that the scalar field should be phantom and that we should change the sign at the kinetic term of the scalar field Lagrangian  \eqref{stat-Lagrange}. Then, instead of Eq. \eqref{stat-scal1}, we have 
\begin{equation}
\phi'^2 = \frac{b^2}{(r^2+b^2)^2}.
\label{stat-scal2}
\end{equation}
Furthermore,
\begin{equation}
\phi' = \pm \frac{b}{r^2+b^2} \,,
\label{stat-scal3}
\end{equation}
and choosing the positive sign in the right-hand side of Eq. \eqref{stat-scal3}, we can integrate it to obtain 
\begin{equation}
\phi = {\rm arctan}\frac{r}{b},
\label{stat-scal4}
\end{equation}
where an integration constant is chosen so that $\phi(0) = 0$. Inversely,
\begin{equation}
r = b \tan \phi.
\label{stat-scal5}
\end{equation} 

Analogously, from  Eqs. \eqref{Ein}, \eqref{stat-en-mom} and \eqref{stat-en-mom1}, we obtain the following expression for the potential of the scalar field 
\begin{equation}
V = \frac12(T_0^0+T_r^r)=\frac12(G_0^6+G_r^r)=\frac12(R_0^0+R_r^r-R) = -R_{\theta}^{\theta}.
\label{stat-scal6}
\end{equation}
Using the expression \eqref{stat-Ricci2} with the condition \eqref{stat-condition2}, we find
\begin{equation}
V = \frac{A-1-rA'}{r^2+b^2}.
\label{stat-scal7}
\end{equation} 
Applying the expression \eqref{stat-exact} for $A$, we obtain 
\begin{eqnarray}
&&V = -\frac{3\pi r_0}{4b^3}+\frac{3r_0}{2b^3}{\rm arctan}\frac{r}{b}+\frac{9r_0r}{2b^2(r^2+b^2)}\nonumber \\
&&-\frac{3\pi r_0r^2}{2b^3(r^2+b^2)}+\frac{3r_0r^2}{b^3(r^2+b^2)}{\rm arctan}\frac{r}{b}.
\label{stat-scal8}
\end{eqnarray}
Using the relation \eqref{stat-scal5}, we can rewrite the expression \eqref{stat-scal8} as an expression for the potential as a function of the phantom scalar field. It has the following form:
\begin{eqnarray}
&&V = -\frac{3\pi r_0}{4b^3}+\frac{3r_0}{2b^3}\phi+\frac{9r_0\sin 2\phi}{4b^3}\nonumber \\
&&-\frac{3\pi r_0\sin^2\phi}{2b^3}+\frac{3r_0\phi\sin^2\phi}{b^3}.
\label{stat-scal9}
\end{eqnarray}
Here, the domain of the function is 
\begin{equation}
0 \leq \phi < \frac{\pi}{2} \,,
\label{domain}
\end{equation}
and the potential is an analytic function without any irregularities of the cusp type. It is in agreement with the fact that, in this solution, the scalar field does not undergo the phantom - non-phantom transition in contrast to the examples considered in the two preceding sections. 

Naturally, there are other regular solutions of the Fisher type sustained by a scalar field. The solution based on the choice of the condition \eqref{stat-condition2} presented here is especially simple. We can say that one can obtain at least another explicit solution if one fixes the function $A$ and solves the equation of the function $B$. However, this solution is much more cumbersome.  

\section{Concluding remarks}
We have applied a procedure for the elimination of singularities, which was proposed in \cite{Simpson-Visser} for black holes, to simple cosmological models. We have seen that the non-singular versions of the flat Friedmann model and the Bianchi I universe, both filled with a homogeneous scalar field, display the transition between the standard scalar field and its phantom counterpart. In this context, this phantom-scalar transition was previously presented in the article \cite{Bronnikov} for the more complicated case of a spherically symmetric static model.

Moreover, a transition between these two types of scalar fields was also investigated in the articles \cite{Vikman:2004,we4,we5} in a rather different context. There, the starting point was the observation that the equation of state of effective dark energy models in the late universe can change its form across the value $w=-1$ (which is sometimes called the ``crossing of the phantom divide line''). Motivated by this fact and by some mathematical observations made in the article \cite{Yurov}, the authors of \cite{we4} proposed a model where this effect is realized in the presence of a single scalar field (see also the earlier work \cite{Vikman:2004}). For this to be achieved in \cite{we4}, it was necessary to have a cusp in the potential of the scalar field, and its initial conditions needed to be chosen in a special way. Further details of this model were explored in \cite{we5}. It is also worthwhile to mention yet another set of works in which a phantom-scalar transition has been analyzed. In \cite{Deffayet:2010}, a transition of this kind was achieved without ghost instabilities. In \cite{Bars, Bars1,bang}, the phantom-scalar transition was combined with a transformation of gravity into anti-gravity (and vice versa) in the process of crossing the big crunch-big bang singularity.

It is clear that the approach of \cite{we4} and \cite{we5} is quite different from that of \cite{Simpson-Visser}, \cite{Bronnikov} and the present article. Furthermore, the physical situations are also distinct: whereas here we examine a regularization of the big crunch-big bang singularity, the late-time accelerating universe was studied in \cite{we4} and \cite{we5}. Nevertheless, the character of the non-analyticity in the potential, which is responsible for the ``(de)-phantomization'' of the scalar field, appears to be the same. It would thus be interesting to further develop the regularized cosmological models presented here in connection with inflation and bouncing models, as well as (effective) quantum gravity. 

\vspace{-0.5cm}
\begin{acknowledgements}
L.C. thanks the Dipartimento di Fisica e Astronomia of the Universit\`{a} di Bologna as well as the I.N.F.N. Sezione di Bologna for financial support.
\end{acknowledgements}
\end{document}